\documentstyle[11pt]{article}
\def\fnote#1#2{\begingroup\def\thefootnote{#1}\footnote{#2}\addtocounter
{footnote}{-1}\endgroup}

\begin{document}

\hfill{UTTG-03-19}

\vspace{20pt}

\begin{center}
{\large {\bf Absorption of Gravitational Waves from Distant Sources}}

\vspace{20pt}

Raphael Flauger\fnote{*}{Electronic address: rflauger@ucsd.edu
}\\
{\em Department of Physics, University of
California, San Diego\\
La Jolla, CA, 92093}

\vspace{20pt}

Steven Weinberg\fnote{**}{Electronic address:
weinberg@physics.utexas.edu}\\
{\em Theory Group, Department of Physics, University of
Texas\\
Austin, TX, 78712}

\end{center}

\vspace{30pt}

\noindent
\begin{center}
{\bf Abstract}
\end{center}

The rate of gravitational wave absorption by inverse bremsstrahlung is calculated.  It increases with decreasing frequency $\nu$ as $\nu^{-3}$.  Nevertheless, because of the near cancellation of absorption by stimulated emission,  the ionized gas in galaxy clusters does not block gravitational waves at the  nanohertz frequencies that may be detected by the use of pulsar timing observations.

\noindent

\vfill

\pagebreak

\begin{center}
{\bf I INTRODUCTION}
\end{center}

The exciting discovery of gravitational waves from coalescing black holes [1] and neutron stars [2] has naturally increased interest [3]  in possible effects that intervening matter may have on these waves.  There is an old result of Hawking [4], that gives the rate of absorption as $\Gamma_{\rm abs}=16\pi G\eta/c^2$, where $\eta$ is the viscosity of the matter, but this only applies if the collision frequency in the matter is much greater than the frequency of the gravitational wave,
which it typically is not.  We have recently studied the opposite extreme case of gravitational wave propagation through collisionless matter [5], but we found no observable effects, except perhaps for cosmological sources.  This paper will consider a case different from either of these: the quantum mechanical absorption of  low frequency gravitational waves during collisions occurring in intervening matter.   

Though we will keep our treatment of this effect as general as possible, in making approximations we  shall have in mind the natural application to clusters of galaxies.  They are big, and contain hot ionized gas with temperatures  of a few keV and electron number densities of $10^{-3}\,{\rm cm}^{-3}$ to $10^{-2}\,{\rm cm}^{-3}$.   We will see that the absorption rate increases sharply with decreasing frequency, so that it might be thought that clusters of galaxies could effectively block gravitational waves with frequency less than a few hundred nanohertz,  just the waves from binaries of supermassive black holes in galaxy clusters that might otherwise be detected by pulsar timing observations [6].  Fortunately, it turns out that this absorption is almost entirely cancelled by stimulated emission,  and does not present an obstacle to this use of pulsar timing observations.

\vspace{20pt}

\begin{center}
{\bf II  ABSORPTION AND EMISSION}
\end{center}

We begin with a reminder of the relation between the rates of emission and absorption of a soft graviton or photon, which will allow us to use an old formula [7] for the rate of gravitational wave emission in collisions  to calculate the  rate of gravitational wave absorption in the same collisions.  Suppose we write the rate of emission of  photons or gravitons with momenta in a range $d^3q$ around ${\bf q}$ in collision processes in a volume $V$ as
\begin{equation}
d\Gamma_{\rm em}=V|{\cal M}|^2d^3q\;,
\end{equation}
where $|{\cal M}|^2$ is proportional to the thermal average of a sum over helicity of a squared matrix element and  products of densities of colliding particles.  Crossing symmetry dictates that the matrix elements for absorption and emission of a very soft neutral massless particle are the same apart from phases, but because in absorption there is one more particle in the initial state the absorption rate has an additional factor $(2\pi\hbar)^3/V$, and of course it does not have the factor $d^3q$ for the final photon or graviton that appears in the emission rate.  Hence the absorption rate is
\begin{equation}
\Gamma_{\rm abs}=\frac{1}{2}(2\pi\hbar)^3|{\cal M}|^2\;,
\end{equation}
with the factor $1/2$ appearing here because in calculating the absorption rate we average rather than sum over helicity.
 The absorption rate can therefore conveniently be expressed in terms of a quantity familiar in astrophysics, the emissivity $j_\nu$, defined as the energy emitted per time, 
 per volume,  per photon or graviton solid angle $d\Omega$, and per frequency interval $d\nu$ for frequencies in the range $\nu$ to $\nu+d\nu$.  Since 
$d^3q=q^2\,dq\,d\Omega=(2\pi\hbar/ c)^3\nu^2d\nu\,d\Omega$, the emissivity is $j_\nu=2\pi\hbar\nu\times(2\pi\hbar/ c)^3 \nu^2 |{\cal M}|^2$, and so the absorption rate is related to the emissivity by
\begin{equation}
\Gamma_{\rm abs}(\nu)=\left(\frac{c^3}{4\pi \hbar\,\nu^3}\right)\,j_\nu\;.
\end{equation}

This result is derived from quantum mechanics alone, without considerations of thermal equilibrium.  It is therefore limited to very low temperature, with 
$kT\ll 2\pi\hbar\nu$, and in particular  does not take into account the effect of stimulated emission.  To deal with the more general case it is easiest to adopt the assumption (not entirely obvious for gravitons) that it is possible to bring the radiation into thermal equilibrium with the medium at temperature $T$.  In this case, the balance between emission and net absorption (subtracting stimulated emission) of photons or gravitons requires [8] that $j_\nu=\Gamma_{\rm net \; abs}(\nu)B(\nu)/4\pi$, where $B(\nu)$ is the black-body energy density per frequency interval
$$
B(\nu)=\frac{16\pi^2\hbar\nu^3}{c^3}\left[e^{2\pi\hbar\nu/kT}-1\right]^{-1}\;.
$$
In place of Eq.~(3), we have then for general temperature
$$
\Gamma_{\rm net\;abs}(\nu)=\left(\frac{c^3}{4\pi \hbar\,\nu^3}\right)\,j_\nu \,\left[e^{2\pi\hbar\nu/kT}-1\right]\;.
$$
This result is derived in an appendix, under the assumption that the particles with which the radiation interacts are in thermal equilibrium with one another, without needing to assume that the radiation itself can be brought into equilibrium with these particles.

The emissivity contains a factor $e^{-2\pi\hbar\nu/kT}$, which combined with the factor $e^{2\pi\hbar\nu/kT}-1$ in Eq.~(4) yields a factor 
$1-e^{-2\pi\hbar\nu/kT}$, with the first and second terms representing the effects of absorption and stimulated emission.  Here we are concerned with the case of very low frequency, for which $2\pi\hbar\nu\ll kT$, so whether or not we take account of the factor $e^{-2\pi\hbar\nu/kT}$ in the emissivity, we have simply
\begin{equation}
\Gamma_{\rm net\;abs}(\nu)=j_\nu \,\left(\frac{c^3}{4\pi \hbar\,\nu^3}\right)\,\left(\frac{2\pi\hbar\nu}{kT}\right)\;.
\end{equation}

\vspace{20pt}

\begin{center}
{\bf III  GRAVITON ABSORPTION}
\end{center}

The rate  for the production of graviton energy $\leq E$ in a single collision of some type  $\alpha\rightarrow \beta$ is given [7] for small $E$ by
\begin{equation}
d\Gamma_{\alpha\rightarrow \beta}(\leq  E)\rightarrow (E/\Lambda)^B\,b(B)\,d\Gamma^0_{\alpha\rightarrow \beta}\;.
\end{equation}
Here
\begin{equation}
B=\frac{G}{2\pi\hbar c}\sum_{n,m}\eta_n\eta_m m_nm_m \frac{1+\beta_{nm}^2}{\beta_{nm}(1-\beta_{nm}^2)^{1/2}}\ln\left(\frac{1+\beta_{nm}}{1-
\beta_{nm}}\right)\;,
\end{equation}
where the sums run over all particles participating in the reaction $\alpha\rightarrow\beta$;  $m_n$  is the rest mass of the $n$th particle; $\eta_n$ equals   $+1$ or $-1$ for  particles in the initial state $\alpha$ or final state $\beta$; and $c\beta_{nm}$  is the  velocity of either of particles $n$ or $m$ in the rest  frame of the other particle:
\begin{equation}
\beta_{nm}\equiv \left[1-\frac{m_n^2m_m^2c^4}{(p_n\cdot p_m)^2}\right]^{1/2}\;.
\end{equation}
Also, $b(B)$ is the function
\begin{equation}
b(B)\equiv \frac{1}{\pi}\int_{-\infty}^{+\infty}  \frac{\sin \sigma\,d\sigma}{\sigma}\exp\left[B\int_0^1\frac{d\omega}{\omega}\Big(e^{i\omega\sigma}-1\Big)\right]=1-\frac{\pi^2B^2}{12}+\dots\;,
\end{equation}
and 
$ d\Gamma^0_{\alpha\rightarrow \beta} $ is the  differential rate 
 for the same process without soft graviton  emission and without radiative corrections from virtual infrared gravitons, where $\Lambda$ is a more-or-less arbitrary upper limit on virtual graviton four-momenta that is used to define what we mean by ``infrared.''  ( $\Lambda$ will not appear in our final results.)  The differential rates $d\Gamma_{\alpha\rightarrow \beta}(\leq E)$ and $d\Gamma^0_{\alpha\rightarrow \beta}$ are rates for producing the particles in the final state $\beta$ in some infinitesimal element of their momentum spaces, the same for both rates.  (The formula given in natural units in ref. [7] has been modified here by inserting a factor  $1/\hbar c$ to make $B$ dimensionless in cgs units.)  Since generally $B\ll 1$. we can approximate $b(B)=1$ and write Eq.~(5) as
\begin{equation}
d\Gamma_{\alpha\rightarrow \beta}(\leq  E)\rightarrow \left[1+B\ln(E/\Lambda)\right]\,d\Gamma^0_{\alpha\rightarrow \beta}\;.
\end{equation}
and the rate of energy radiation per collision and per frequency interval at low frequency is then 
\begin{equation}
2\pi\hbar\nu\frac{d}{d\nu}d\Gamma_{\alpha\rightarrow \beta}(\leq  2\pi\hbar\nu)\rightarrow 2\pi\hbar B\,d\Gamma^0_{\alpha\rightarrow \beta}\;.
\end{equation}

Formula (10)  applies for relativistic or non-relativistic processes involving any number of particles of arbitrary spin, whatever the interactions may be that produce the reaction $\alpha\rightarrow\beta$.  If we now specialize to the case where $\alpha\rightarrow \beta$ is    non-relativistic  elastic two-body scattering, and take into account the conservation of energy and momentum, we have [7]
\begin{equation}
B=\frac{8G}{5\pi \hbar c^5}\mu^2 v^4\sin^2\theta_c\;.
\end{equation}
where $\mu$ is the reduced mass, $v\equiv |{\bf v}_1-{\bf v}_2|$ is the relative speed, 
 and $\theta_c$ is the scattering angle in the center-of-mass system.  
 Eqs.~(10) and (11) then give the emissivity at low frequency as
\begin{equation}
j_\nu\rightarrow\frac{2\pi\hbar}{4\pi}\frac{8G\mu^2}{5\pi \hbar c^5}n_1n_2\,\overline{v^5\sigma_D}\;,
\end{equation}
where $n_1$ and $n_2$ are the number densities of the two colliding particles, $\sigma_D$ is a deflection cross section
\begin{equation}
\sigma_D\equiv \int \frac{d\sigma}{d\Omega'}\sin^2\theta_c d\Omega'\;,
\end{equation}
and the bar indicates an average over incident velocities.

Using Eq.~(4) now gives a general formula for the net rate at which gravitational waves of low frequency $\nu$ with $2\pi\hbar\nu\ll kT$ are absorbed in non-relativistic two-body collisions
\begin{equation}
\Gamma_{\rm net\;abs}(\nu)\rightarrow \frac{G\mu^2}{5\pi^2\hbar c^2\nu^3}n_1n_2\;\overline{v^5\sigma_D}\left[\frac{2\pi\hbar\nu}{kT}\right]\;.
 \end{equation}
the final factor $2\pi\hbar\nu/kT$ arising from the near cancellation of absorption by stimulated emission.

We now specialize further, and consider the Coulomb scattering of electrons by protons in fully ionized hydrogen.  Here $\mu$ is close to the electron mass $m_e$; $v$ is close to the initial electron velocity; $n_1=n_2$ is the electron number density $n_e$, and the cross-section (13) is
\begin{equation}
\sigma_D=\frac{2\pi e^4}{m_e^2 v^4}\int_{-1}^{+1} \frac{\sin^2\theta \;d\cos\theta}{[1-\cos\theta+\hbar^2/2m_2^2v^2\ell^2]^2}\;,
\end{equation}
 where $e$ is the electron charge in unrationalized electrostatic units, and $\ell\equiv \sqrt{kT/4\pi n_e e^2}$ is the Debye screening length.  Eq.~(15) is derived using the Born approximation, which applies in intracluster gas because  typical electron kinetic energies are much larger than a Rydberg, 13.6 eV.  Also, in Eq.~(15)   we are neglecting the difference $2\pi \hbar\nu$ between initial and final electron energies, which is a good approximation because we are concerned with gravitational wave frequencies much less than the plasma frequency, which in intracluster gas is a few hundred Hz.  

As we shall see, in the cases that interest us here $m_ev\ell/\hbar$ is very large for typical values of $v$, and Eq.~(15) therefore gives
\begin{equation}
\sigma_D=\frac{4\pi e^4}{m_e^2v^4}\left[-2+\ln\left(\frac{4m_e^2v^2\ell^2}{\hbar^2}\right)\right]\;.
\end{equation}
The thermal average in Eq.~(14) is then, for $2\pi\hbar \nu\ll kT$, 
\begin{equation}
\overline{v^5\sigma_D}=\frac{8 e^4(2\pi kT)^{1/2}}{m_e^{5/2}}\left[\ln\left(\frac{8m_e\ell^2kT}{\hbar^2}\right)-1-\gamma\right]\;,
\end{equation}
where $\gamma=0.577\dots$ is the Euler constant.

Before drawing consequences from Eqs.~(14) and (17), we need to check  that the absorption occurs in independent collisions, rather than in an imperfect fluid as was assumed in ref. [4].  So we need to ask, what is the rate $\nu_C$ of relevant collisions?    The absorption of gravitons in a collision of an electron with a proton is unaffected if at the same time the electron experiences forward scattering by the Coulomb field of some distant other proton, so the cross section to use 
in estimating $\nu_C$ is not the total cross section, but something like the deflection cross section $\sigma_D$.  Also, we are not interested in  collisions of very slow electrons which,  because  the factor $v^4$ in Eq.~(11) cancels the factor $v^{-4}$ in Eq.~(15), contribute little to gravitational wave emission or absorption.  Therefore instead of taking $\nu_C$ as the thermal average of $n_ev\sigma_D$, we will take it as the average weighted with an additional factor $v^4$:
\begin{equation}
\nu_C=n_e\overline {v^5\sigma_D}/\overline{v^4}\;,
\end{equation}
where the bar again indicates an ordinary thermal average.  (It would make little difference numerically if we weighted the average over velocity with any power $v^n$ with $n\geq 1$ instead of $v^4$.) 
The argument of the logarithm is 
\begin{equation}
\frac{8m_e\ell^2kT}{\hbar^2}=\frac{2m_e(kT)^2}{\pi e^2n_e\hbar^2}= 6\times 10^{30} \Big[kT ({\rm keV})\Big]^2 \Big[n_e (10^{-3} {\rm cm}^{-3})\Big]^{-1}\;.
\end{equation}
This is so large that changes of a few orders of magnitude in $kT$ or $n_e$ make little difference in the logarithm, so we shall fix the quantity in square brackets in Eq.~(17) to have the value  $\ln (6\times 10^{30})-1-\gamma=69.3$.  
The effective collision frequency (18) is then
\begin{eqnarray}
&&\nu_C=\frac{4}{15}\left(\frac{m_e}{2kT}\right)^2\times \frac{8 n_e e^4(2\pi kT)^{1/2}}{m_e^{5/2}}\left[\ln\left(\frac{8m_e\ell^2kT}{\hbar^2}\right)-1-\gamma\right]\nonumber\\&&
= 2.5\times 10^{-12} {\rm sec}^{-1} \Big[n_e(10^{-3}\,{\rm cm}^{-3})\Big]\Big[kT({\rm keV})\Big]^{-3/2} \;.
\end{eqnarray}

There is still a question, whether the appropriate condition that allows us to treat the collisions in which gravitons are absorbed as independent is that electrons experience many cycles of the gravitational wave between collisions, which requires that $\nu\gg \nu_C$, or that electrons  pass through many gravitational wavelengths between collisions, which requires that the mean free path $v/\nu_C$ be much longer than the wavelength $c/\nu$ for typical electron velocities $v$, or in other words, that $\nu\gg (c/v)\nu_C$ .  Since $v<c$ the second condition is always more stringent.  For $kT\simeq 1$ keV electrons typically have 
$v/c\simeq 1/30$, so for relevant temperatures and densities $(c/v)\nu_C $  is sufficiently less than the gravitational wave frequencies we will consider so that we can use Eq.~(14) for the graviton absorption rate.

Eqs.~(14) and (17) now give the net absorption rate for $2\pi \hbar \nu\ll kT$:
\begin{eqnarray}
&&\Gamma_{\rm net\;abs}(\nu)=\frac{Gm_e^2}{5\pi^2\hbar c^2\nu^3}n_e^2\times \frac{8 e^4(2\pi kT)^{1/2}}{m_e^{5/2}}\left[\ln\left(\frac{8m_e\ell^2kT}{\hbar^2}\right)-1-\gamma\right]\Big(2\pi\hbar\nu/kT\Big) \nonumber\\&&=\frac{1.4  \times 10^{-34} {\rm sec}^{-1} \Big[n_e(10^{-3}{\rm cm}^{-3})\Big]^2 \Big[kT({\rm keV})\Big]^{1/2}}{\Big[\nu({\rm sec}^{-1})\Big]^3}\Big(2\pi\hbar\nu/kT\Big)\;.
\end{eqnarray}
If it were not for the cancellation of absorption by stimulated emission, represented by the final factor $2\pi\hbar\nu/kT$, the mean distance $c/\Gamma_{\rm abs}$ for graviton absorption  in fully ionized hydrogen with density $n_e\simeq 10^{-3} {\rm cm}^{-3}$ and temperature $kT\simeq 1 $ keV would be less than 1 Mpc at frequencies less than  240 nanohertz.  This covers the range of frequencies of gravitational waves that might be detected by observation of pulsar timing [6].  Fortunately, for $kT\approx 1\,$keV and $\nu\approx 200$ nanohertz the net absorption is suppressed by the factor $2\pi\hbar\nu/kT\approx 10^{-24}$, and has no relevant effect on gravitational wave propagation.  Indeed, if nanohertz gravitational waves are observed coming from galaxy clusters, it will show that gravitons like photons are produced by stimulated emission.

Since gravitational interactions are universal, gravitational waves may also be absorbed in intergalactic space in collisions other than the electron-proton collisions considered here, such as collisions of possible dark matter particles that interact strongly with one another.
The net absorption rate in any non-relativistic elastic two-body  collisions may be calculated using Eq.~(14), or in more general collisions by using Eqs.~(4), (10), and (6).  
\vspace{10pt}

\begin{center}

{\bf Appendix}

\end{center}

In this appendix we will derive a general formula for the rate of change of the occupation number $n({\bf q},\lambda) $ of gravitons or photons interacting with a hot gas, that will exhibit the effects of stimulated emission as well as absorption and spontaneous emission.    The occupation number is defined so that $n({\bf q},\lambda)d^3q/(2\pi\hbar)^3$ is the number density of gravitons or photons of helicity $\lambda$ in a volume $d^3q$ of momentum space around momentum ${\bf q}$.  Its rate of change due to absorption and spontaneous and stimulated  emission  in collisions of particles $1, 2, \dots$ is
\begin{eqnarray}
&&\frac{dn({\bf q},\lambda)}{dt}=-\frac{n({\bf q},\lambda)}{2\pi \hbar}\int d^3p_1\;d^3p_2\cdots d^3p'_1\;d^3p'_2\cdots n_1({\bf p}_1)n_2({\bf p}_2)\cdots \nonumber\\&& \times\delta^3({\bf q}+{\bf p}_1+{\bf p}_2+\dots-   {\bf p}'_1-{\bf p}'_2-\dots)  
\delta(|{\bf q}|c+E_1+E_2+\dots-E_1'-E_2'-\dots)\nonumber\\&&~~~~~\times
\Big|M(\lambda, {\bf q},{\bf p}_1,{\bf p}_2\dots\rightarrow   {\bf p}'_1,{\bf p}'_2\dots)\Big|^2\nonumber\\&&
+\frac{1+n({\bf q},\lambda)}{2\pi \hbar}\int d^3p_1\;d^3p_2\cdots d^3p'_1\;d^3p'_2\cdots n_1({\bf p}_1)n_2({\bf p}_2)\cdots \nonumber\\&& \times\delta^3({\bf p}_1+{\bf p}_2+\dots- {\bf q}-  {\bf p}'_1-{\bf p}'_2-\dots)
\delta(E_1+E_2+\dots-|{\bf q}|c-E_1'-E_2'-\dots)\nonumber\\&&~~~~~\times
\Big|M({\bf p}_1,{\bf p}_2\dots\rightarrow \lambda, {\bf q}, {\bf p}'_1,{\bf p}'_2\dots)\Big|^2\;.
 \end{eqnarray}
Here $M$ is the coefficient of the energy and momentum conservation delta functions in the S-matrix element for the indicated process, and $n_1$, $n_2$, {\em etc.} are the occupation numbers for the colliding particles.  We assume that $n_1$, $n_2$, {\em etc.} are all much less than unity, so that we do not need to take account of the Pauli exclusion principle where the colliding particles are fermions, or of the stimulated emission of the colliding particles if they are bosons.  As usual, the factor $1+n({\bf q})$  arises from the commutator of the photon or graviton creation operator with the $n+1$ annihilation operators in the adjoint of the final state.  The matrix element $M$ and the occupation numbers of the  colliding particles depend on spin indices, which are suppressed; they are understood to be summed along with integrations over momenta.   
 
We now interchange the labels of the momenta (and spins) of the colliding particles in the first term of Eq.~(22),  and make use the unitarity of the S-matrix, which implies that for any multiparticle transition
\begin{equation}
\int d\beta\; \delta^3({\bf p}_\beta-{\bf p}_\alpha)\delta(E_\beta-E_\alpha)\Big|M(\beta\rightarrow \alpha)\Big|^2
=\int d\beta \;\delta^3({\bf p}_\beta-{\bf p}_\alpha)\delta(E_\beta-E_\alpha)\Big|M(\alpha\rightarrow \beta)\Big|^2\;,
\end{equation}
where  $\int d\beta$ is understood to include a sum over the spins of all particles in the state $\beta$ as well as an integration over all 3-momenta of these particles.
We then have
\begin{eqnarray}
&&\frac{dn({\bf q},\lambda)}{dt}=\frac{1}{2\pi \hbar}
\int d^3p_1\;d^3p_2\cdots d^3p'_1\;d^3p'_2\cdots \nonumber\\&&\times \delta^3({\bf p}_1+{\bf p}_2+\dots- {\bf q}-  {\bf p}'_1-{\bf p}'_2-\dots)
\delta(E_1+E_2+\dots-|{\bf q}|c-E_1'-E_2'-\dots)\nonumber\\&&\times
\Big|M({\bf p}_1,{\bf p}_2\dots\rightarrow \lambda, {\bf q}, {\bf p}'_1,{\bf p}'_2\dots)\Big|^2
\nonumber\\&&~~~~~\times\left[-n({\bf q},\lambda)n_1({\bf p}'_1)n_2({\bf p}'_2)\cdots
+\Big(1+n({\bf q},\lambda)\Big)n_1({\bf p}_1)n_2({\bf p}_2)\cdots\right]\;.
\end{eqnarray}

We now assume that the colliding particles are in thermal equilibrium with each other, though not necessarily with the photons or gravitons.  Since we are assuming that their occupation numbers are small, for 
$E_1+E_2+\dots=|{\bf q}|c+E_1'+E_2'+\dots$ we have
\begin{equation}
\frac{n_1({\bf p}_1)n_2({\bf p}_2)\cdots
}{n_1({\bf p}'_1)n_2({\bf p}'_2)\cdots}=\exp\Big(-|{\bf q}|c/kT\Big)=\exp\Big(-2\pi\hbar\nu/kT\Big)
\end{equation}
so
\begin{eqnarray}
&&\frac{dn({\bf q},\lambda)}{dt}=\frac{1}{2\pi \hbar}
\int d^3p_1\;d^3p_2\cdots d^3p'_1\;d^3p'_2\cdots \nonumber\\&&\times \delta^3({\bf p}_1+{\bf p}_2+\dots- {\bf q}-  {\bf p}'_1-{\bf p}'_2-\dots)
\delta(E_1+E_2+\dots-|{\bf q}|c-E_1'-E_2'-\dots)\nonumber\\&&\times
\Big|M({\bf p}_1,{\bf p}_2\dots\rightarrow \lambda, {\bf q}, {\bf p}'_1,{\bf p}'_2\dots)\Big|^2\,n_1({\bf p}_1)n_2({\bf p}_2)\cdots
\nonumber\\&&~~~~~\times \left[1-n({\bf q},\lambda)\Big[\exp\Big(2\pi\hbar\nu/kT\Big)-1\Big]\right]\;.
\end{eqnarray}
The first term $+1$ in the square brackets on the last line arises from spontaneous emission, while the terms  
$-\exp\Big(2\pi\hbar\nu/kT\Big)$ and $+1$ multiplying $n({\bf q},\lambda)$ arise respectively from absorption and stimulated emission.  The important point for this paper is that the ratio of stimulated emission to absorption is $\exp\Big(-2\pi\hbar\nu/kT\Big)$.

\vspace{15pt}

We are grateful to Aaron Zimmerman for helpful conversations about pulsar timing observations and other matters.  This article is based on work of R. F.  supported in part by the Alfred P. Sloan Foundation, the Department of Energy under grant de-sc0009919, and the Simons Foundation/SFARI 560536. and of S. W. supported by the National Science Foundation under Grant Number PHY-1620610, and with support from the Robert A. Welch Foundation, Grant No. F-0014.

\vspace{20pt}

 \begin{center}
REFERENCES
\end{center}
\begin{enumerate}
  
\item B. F. Abbott {\em et at.} [LIGO and Virgo collaborations], Phys. Rev. Lett. {\bf 116}, 061102 (2016) [arXiv: 1602.03837].

\item B. P Abbott {\em et al.}    (LIGO Scientific  Collaboration, Virgo Collaboration, and other collaborations), Astrophys. J. {\bf 848}, L12 (2017).

\item For  instance, see E. Calabrese, N. Battaglia, and D. N. Spergel, Class. Quant. Grav. {\bf 33}, 165004 (2016).

\item S. W. Hawking, Ap. J. {\bf 145}, 544 (1966). 

\item R. Flauger and S. Weinberg, Phys. Rev. D {\bf 75}, 123505 (2007).  

\item S. Detweiler, Ap. J. {\bf 234}, 1100 (1979).  For a current review, see S. Burke-Spolaor {\em et al.}, astro-ph/1811.08826..

\item S. Weinberg,   Phys. Rev.  {\bf 140}, B516 (1965). 
 
\item For instance, see F. H. Shu, {\em The Physics of Astrophysics}, Vol. 1 (University Science Books, Mill Valley, CA, 1991).
  
\end{enumerate}

\end{document}